\documentstyle[multicol,aps,epsfig]{revtex}

\newcommand{\be}{\begin{equation}}
\newcommand{\bel}[1]{\begin{equation}\label{#1}}
\newcommand{\ee}{\end{equation}}
\newcommand{\bea}{\begin{eqnarray}}
\newcommand{\ba}{\begin{array}}
\newcommand{\eea}{\end{eqnarray}}
\newcommand{\ea}{\end{array}}

\begin{document}

\twocolumn[\hsize\textwidth\columnwidth\hsize\csname@twocolumnfalse%
\endcsname

\title{Optimized Traffic Flow at a Single Urban Crossroads : Dynamical
Symmetry Breaking }
%                              %
\author{M. Ebrahim Fouladvand$^{1,3}$ , Nima Hamedani Radja $^{2,3}$  }

\address{$1$ Department of Physics, Zanjan University, P.O.
Box 313, Zanjan, Iran.}

\address{$2$ Department of physics, Sharif university of technology, P.O.
Box 11365-9161 ,
Tehran,Iran. }

\address{$3$ Institute for Studies in Theoretical Physics and Mathematics
,P.O. Box 19395-5531, Tehran, Iran.}

\date{\today}

\maketitle

\begin{abstract}

We propose a stochastic model for the intersection of two urban streets.
The traffic state at the crossroads is controlled by a set of traffic
lights which periodically switch to red and green with a total period
of $T$. Vehicular dynamics is simulated and total delay experienced by
the traffic is evaluated within a definite time interval. Minimising the
delay
give rises to the optimum signalisation of traffic lights. It is shown
that
two different traffic phases are identified in the symmetrically loaded
crossroads. In the light traffic phase, the green times should
be divided  to the streets on an equal basis. In the congested phase,
Minimising the delay enforces us
to break the symmetry between the roads by allocating the majority of
green time to one of the roads. In contrast to 
common sense, equally dividing the green times to the roads, leads to
unfairness to drivers.

\end{abstract}
]
%%%%%%%%%%%%%%%%%%%%%%%%%%%%%%%%%%%%%%%%%%%%%%%%%%%%%%%%%%%%%%%%%%%%%%%%

\section{Introduction}

Modelled as a system of interacting particles driven far from equilibrium,
{\it vehicular traffic } provides the possibility to study various aspects
of truly non-equilibrium systems which are of current interest in
statistical physics \cite{css99,helbbook,tgf97,tgf99}. For almost
half
century, physicists have been challenged to understand the fundamental
principles governing the vehicular flow
\cite{tgf97,tgf99,prigogine}.
Recently, discrete models such as {\it cellular automata } have
provided a significant theoretical framework for modelling traffic flow.
The first cellular automata was introduced by {\it Biham, Middleton
} and {\it Levine}-known as the so-called BML model- which described a
simplified network of urban
crossroads \cite{bml}. soon after, cellular automata, found their way in
highway
traffic through the pioneer work of {\it Nagel} and {\it Schreckenberg}
\cite{ns92} which became the ancestor of many papers in
the literature ( for a review see ref \cite{css99} and the references
therein ). The BML model itself was later generalized to take into
account several
realistic features such as faulty traffic lights \cite{chung}, independent
turning of vehicles \cite{cuesta,nagatani}, and green-wave synchronization
\cite{torok}. The Nagel-Schreckenberg and BML model were recently combined
to cast an upgraded version of urban network models \cite{cs}. In a very
recent paper, the
model is now extended to account for different types of global
signalisation
\cite{brockfeld}.\\
Despite these efforts and those carried out by traffic
engineers, the question of an optimal signalisation scheme for a realistic
urban network
has not yet been comprehensively reviewed.
In the above approaches, the main concern has been focused on the
global strategies of the traffic network and frequently the role of
isolated crossroads have been suppressed. We believe that the optimisation
of
traffic flow at a single crossroads is a substantial ingredient towards a
global optimisation. Isolated crossings are fundamental operating units
of the sophisticated and correlated urban network and thorough analysis of
them would be advantageous toward the ultimate task of the global
optimisation
of the city network. In this regard, our objective in this paper is to
analyse the traffic state of an isolated crossroads in order to find a
better insight into the problem. In addition to theoretical viewpoint,
an investigation of isolated crossroads could be of practical importance.
To
a very good approximation, marginal crossroads in cities are unaffected by
other crossings and can be regarded as isolated ones. There are two basic
types of control for traffic lights at intersections: {\it fixed-cycle}
and {\it traffic-responsive}. Both of these methods can be implemented via 
centralized or decentralized strategies. The application of each
method strongly depends on traffic condition and the topology of the city
network. In this paper we study the impact of randomness in  vehicular
demand at a single crossroads which is controlled under a decentralized
fixed-cycle scheme.

\section{ Formulation of the Model}

An isolated crossroads is formed at the intersection of two streets. These
streets, in principle, can each carry two oppositely flows of vehicles.
Depending on the designing of the crossroads, different phases of
movement can be defined ( a phase of traffic is defined as the flow of
vehicles that proceed a crossroads without conflict). Here for simplicity
we restrict ourselves
to the simplest structure of a crossroads: a {\it one-way to one-way }
intersection. With no loss of generality, we take the direction of
the flow in the first street, hereafter referred to as the
street A, northwards.  The other street ( hearafter referred to as
street B) conducts a one-way eastward flow. Cars arrive at the south and
west entrances of the crossroads. The traffic flow is controlled by a set
of traffic lights. We assume that the traffic is controlled by a
fixed-cycle scheme. In this scheme, the lights periodically go green with
a fixed period of $T$. This period is divided into two parts : in the
first part, the traffic light is green for street $A$ and
simultaneously red for street $B$. This part lasts for $T_{green}$
seconds ( $ T_{green} < T $). In the second part, the lights change
colour and the movement is allowed for the cars of road B. The second
part lasts from $T_{green}$ to $T$. This picture is repeated
periodically. Cars enter the the crossroads and a fraction of them
experience the red light and consequently have to wait until they are
allowed to
leave the intersection during the upcoming green period. The basic
question
 is\\

{\it how to
adjust the ratio of $\frac{T_{green}}{T}$ in order to optimise the
throughput flow?}.\\

There is now an almost well-established agreement on
the
quantitative definition of optimisation. Borrowing from the traffic
engineering literature, we adopt {\it optimised traffic } as
 a state
in which the total delay of vehicles is minimum. In one of our earlier
works \cite{foolad}, we analytically evaluated the total delay in terms of
arrival as well as that of the 
exit rates of vehicles. However, our approached was based on
the simple assumption of the time-constancy of the arrival rates. In
reality,
we know
that successive cars arrive with fluctuating time-headways which
consequently induces 
time-varying arrival rates. In this paper we address the question of
non-constant rates. In order to evaluate the delay, we have simulated the
flow of vehicles at the crossroads. For the sake of
simplicity, we have assumed that each street has a single lane. For
streets with more than one lane, one simply should multiply the value
of delay by the number of lanes. Moreover, we allow those fractions of
northward ( eastward) cars tending
to
turn right (
left) manage to turn via a by-pass road, therefore the cars in our
simulation denote those which wish to go through the crossroads without
turning.
In the next sub section we state our dynamical rules for vehicle
movement.

\subsection { Vehicular dynamics at the crossroads }
          
Our crossroads is modelled by a set of two perpendicular chains.
Each chain is divided into cells which are the same size as a typical
car length . We take the number of cells to be $L$ for both roads. Each
cell can be either occupied by a car or being empty. A traffic light
periodically changes from red to green. The period of the light is taken
to be $T$ units of time and remains constant. The system evolves under a
discrete-time
dynamics. At each step of time, the system is characterized by the
vehicular configuration at each road and the traffic light state of road
A (north-bound vehicles). The state of the system at time $t+1$ is
updated from that in time $t$ by applying the following rules.\\

{\bf step 1 : signal determination}.\\

We first specify the signal state. If the remainder of division of
$t+1$ by $T$ is less than $T_{green}$, then the signal state is defined as
being green for street A and red for street B. Otherwise we set the signal
red
for street A and street for road B.\\

{\bf step 2 : movement in the green road}.\\

At this stage, we update the position of cars on the green road. This step
is divided into two sub-steps. Denoting
the position of the $i$-th cell at time $t$ by  pos[i,t], the following
half-step rule updates the position of cars synchronously ($i=1$ is the 
nearest site to the crossroads).\\
\be
pos[i,t+{1\over 2}] = pos[i+1,t]  ~~~ i=1 \cdots L-1 
\ee

In the above formula, $pos[j,t]=1$ if the site $j$ is occupied and zero
otherwise. According to the above rule, each car moves deterministically 
one cell forward.
This type of parallel movement specifies a uniform displacement of cars, 
i.e.,
the velocities of all cars are taken to be equal. Nevertheless,
in reality,
those cars which observe an empty spatial gap may accelerate to fill the
gap and increase their chance of going through the green light. To
implement 
this very fact in our model, we have added a probabilistic sub-step to
the previous deterministic sub-step. A car with a space-headway
greater than
zero, move another extra cell forward with the probability
$p_{acc}$ according to the following rule.
\be
if~~ pos[i,t+{1 \over 2}]=1 ~~and~~ pos[i-1,t+{1 \over 2}]=0 
\ee
then with the probability $p_{acc}$ the following rule occurs.
\be
pos[i,t+1]:=0 ;~~~~ pos[i-1,t+1]:=1
\ee
In addition, we have assumed that in the green period, no car  
waits and therefore does not contribute to the delay.\\

{\bf step 3 : movement of red road, delay evaluation }.\\

Similar to the previous step, the updating is divided into two
sub-steps. In the first half, we evaluate the delay of cars waiting on the
red period of the signalisation. In the second half, we update the
position
of the moving cars approaching toward the waiting queue. We should note
that
once the signal switches to red,
the moving cars continue their movement until they come to a complete
stop upon
reaching to the end of the waiting queue. As soon as a car comes to halt,
it
contributes to the total delay. In order to evaluate the delay, we measure
the queue length (the number of stopped cars) at time step t and
denote it by the variable $Q$. We remind that\\
\be
pos[i,t]=1~for~ i=1 \cdots Q ~and~ 0~ for
~j=Q+1 \cdots L
\ee
Delay at time step $t+1$ is obtained by adding the queue length $Q$ to the
delay at time step $t$.
\be
delay(t+1)= delay(t)+ Q(t)
\ee

This ensures that during the next time step, all the
stopped cars contribute one step of time to the delay. The next sub-step
describes the position update of moving cars. Moving cars can
potentially be
found in the
cells $ Q+1,Q+2, \cdots, L$.  These cars will move one step forward. The
possibility of double hopping has been rejected since we have assumed that
once a driver notifies the red signal, (s)he no longer attempts to
accelerates.\\  

{\bf step 4 : entrance of cars to the crossroads}.\\

So far, we have dealt with those cars within the horizon of the 
crossroads (site $L$).
Now we let the cars enter into this horizon. To fulfill this
task, at the
end of the movement rules, we evaluate the position of most remote cars on
both streets. We denote them by $last_A$ and $last_B$. By definition ,
$pos_A[j,t]=0$
for $j>last_A$.  A similar statement applies to street B. From our
everyday
driving experience, we empirically observe that the time head ways between 
cars randomly varies which consequently implies a random distance headway
between successive cars. In order to simulate the entrance behaviour of
cars into the horizon of the crossroads, we choose a random integer number
between zero and nine. These numbers, representing the distance-headway
of the oncoming successor of the farthest car, should be chosen from a
realistic
random distribution function. Let us denote these headways by $h_A$ and
$h_B$ for 
street A and B respectively. In our simulation, we have tested Gaussian
distribution function with a wide range of average as well as standard
deviation. Once the random distance headway is chosen, we create a car at
the position  $last_A+h_A$ ($last_B+h_B$) of the street A (B)
respectively.
The
car creation procedure is valid provided the following constraints are
satisfied:
\be
last_A +h_A \leq L ~~$ and ~~$ last_B+h_B\leq L 
\ee

If the position of the created car exceeds the horizon length $L$, the
creation is rejected. The above {\it ad hoc} rules updates the
configuration of the crossroads in the next time step.
before turning to our results, some points worth mentioning. First, the
movement
rules of our model differ quantitatively from the existing one in the
{\it highway traffic literature}. In our model, the natural time step is
chosen to be
the time required for a typical car to replace its predecessor. In
contrast
to the car-following model of Nagel-Schreckenberg where cars can
change their velocity, we have assumed the cars' velocities is almost
constant for all vehicles. This assumption is justified by the empirically
observed fact that velocity fluctuation is suppressed within the horizon
of
the crossroads. The situation may differ drastically in high way
traffic where one observes a considerable fluctuation in the velocity of
cars. Every day measuring of cars velocity in urban areas- where 
speed-limit regulations e.g. fifty kilometres per hour should be
observed by the drivers -supports our
assumption of a
uniform velocity. The other point to mention concerns the total delay.
According to the above rules, no contribution has been devoted to the
green phases and we ideally have assumed upon switching the signal from
red to green, the waiting cars instantaneously reach to the asymptotic
velocity. In
reality, however, it takes a few seconds for the waiting queue to react to
the
signal change. In our model, we have ignored this off-reaction
contribution. In the next section, we will explain our simulation results.

\section { Simulation results : dynamical symmetry breaking }

\subsection { simulation parameters}

Before expressing our main results, let us specify the numeric values of
the
model parameters. The horizon length of the crossroads have been equally
taken to be 100 cells for both roads. Setting the bumper-to-bumper
distance ( 7 metres) between successive cars as the cell length, the real
horizon length is estimated to be 700 metres. Beyond this distance, it is
assumed
that drivers do not control the signals and that driving strategy is not
affected by
the traffic light state. The probability of double hopping is taken to be
$p_{acc}=0.4$. We let the crossroads evolve for 5000 time steps which
is approximately equal to a real two-hour period. Our data has been
averaged over 50 run of the programme.
We let the green time of street A vary from zero to $T$. For each value
of
$T_{green}$, we evaluate the total delay for both streets as well the
number
of passed vehicle during 5000 time steps. 

\subsection{symmetric state: light flow}

Here the traffic conditions are equal for both roads. First we
discuss the light traffic state. In this case, we equally load the
crossroads
with entering cars temporally separated by random time-headways from each
other. The
following graphs depicts the total delay as a function of $T_{green}$
allocated to road A. For the light traffic state which is characterized by
large time headway, we observe a flat time interval inside which
the total delay remains almost constant. Near the boundary of this area,
there is a discontinuous increment of total delay which can be interpreted
as a kind of dynamical phase transition. These results can lead to a
feasible application by
traffic engineers. According to our simulation results, we can 
set the $T_{green}$ at ${T \over 2}$. The existence of an optimum time
interval gives us a desirable freedom to account for the probable
effect of entrance rate variations. As can be seen from the graphs,
increasing the average time-headway leads to a broader optimum interval.

\begin{figure}\label{Fig1}
\epsfxsize=5.5truecm
\centerline{\epsfbox{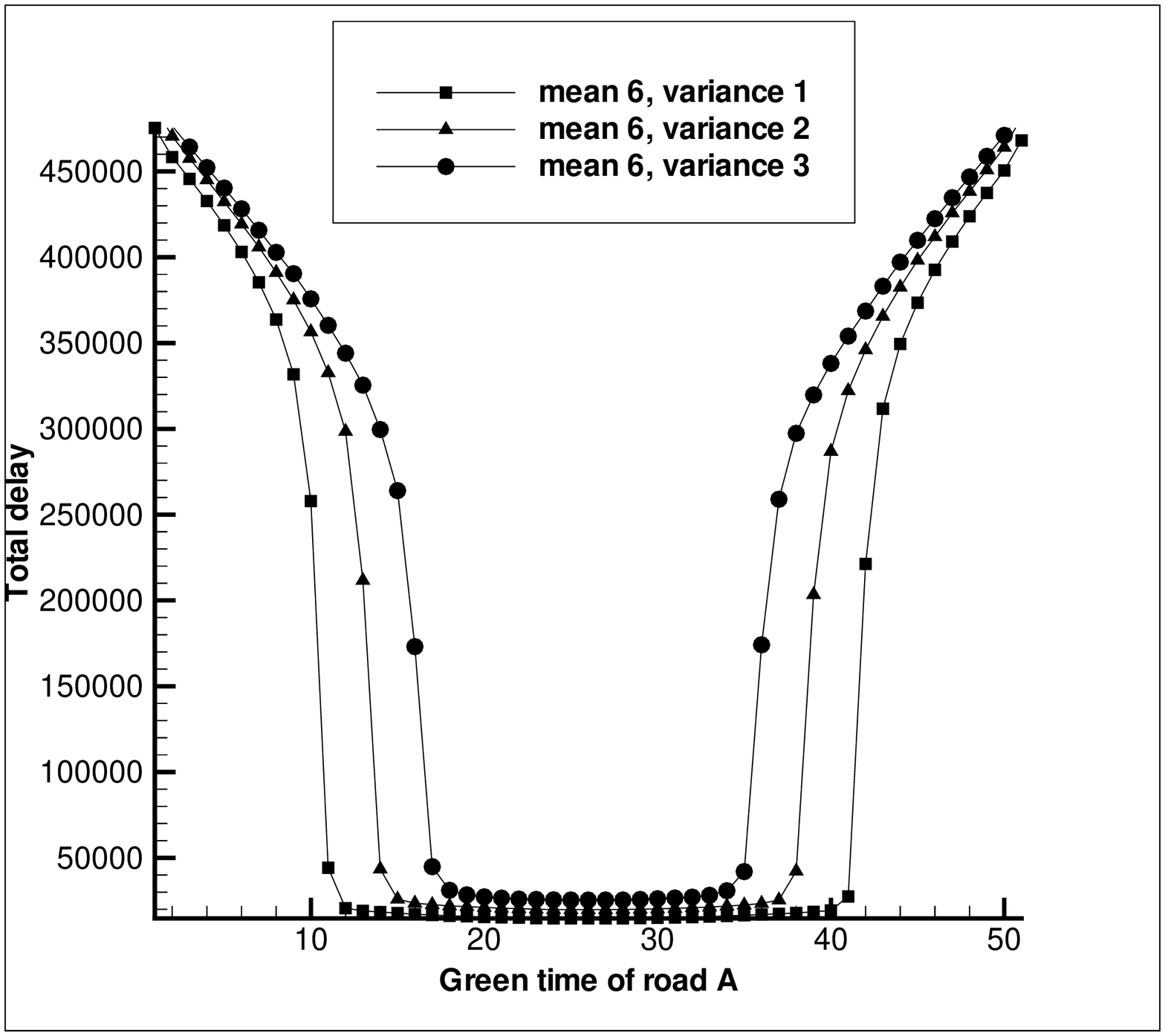}}
\end{figure}
\vspace{0.05 cm}
{\small{Fig.~1: a symmetric light state. Total delay versus the green
time of road A is sketched. As depicted,
for fixed average, increasing the variance leads to broadening of the flat 
 region. $T$ is set to 50 and the crossroads has been evolved for 5000
time-step.} }

\begin{figure}\label{Fig2}
\epsfxsize=5.5truecm
\centerline{\epsfbox{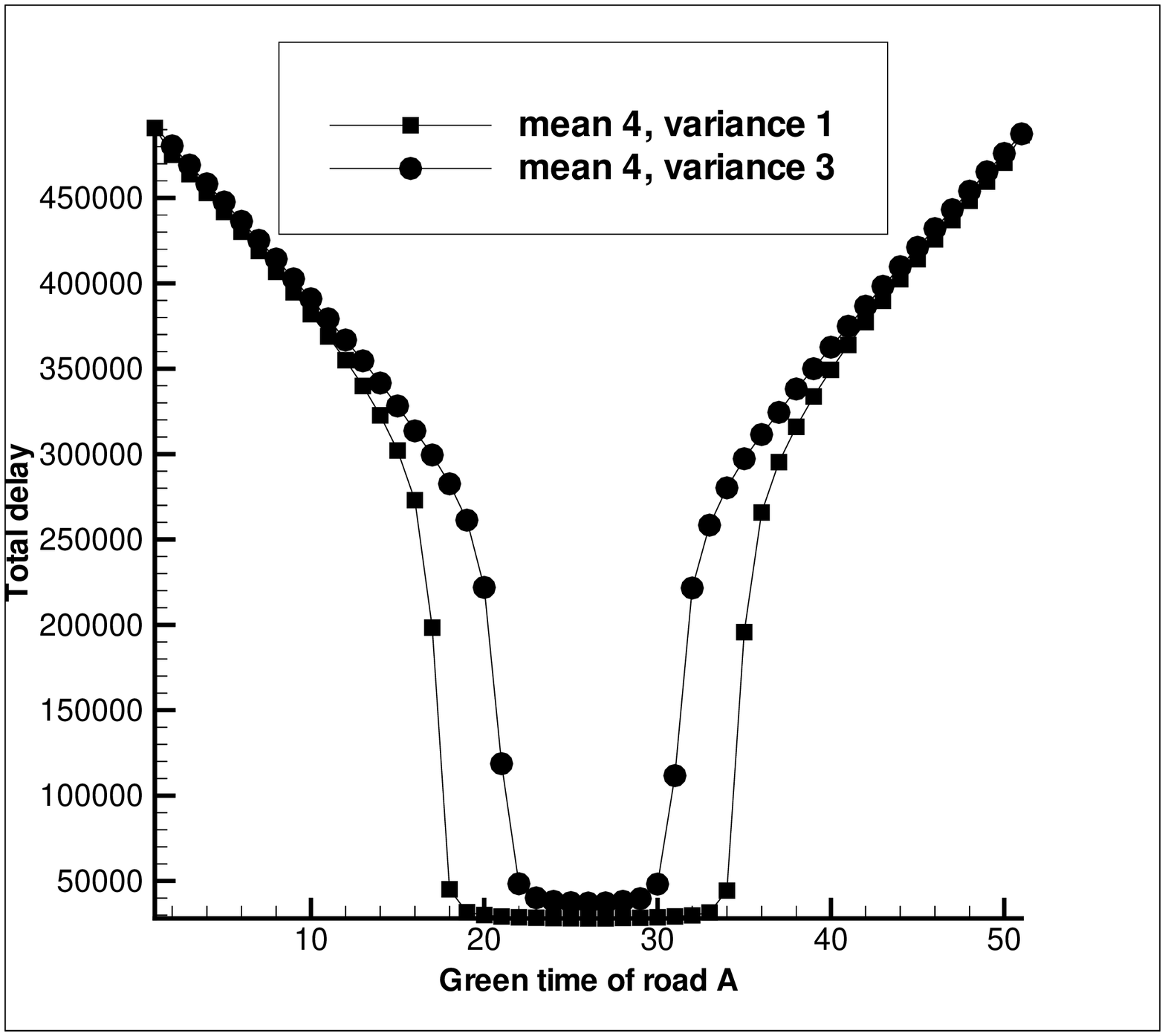}}
\end{figure}
\vspace{0.05 cm}
{\small{Fig.~2: a more congested state with an average space gap of four
cells. The flat region has been shortened in comparison to above graph. The
value of $T$ is 50 and the system has evolved for 5000 time-steps. } }

\subsection{ congested flow : symmetry broken phase }   

Perhaps the most noticeable result of this paper lies in the congested
phase
of the traffic. For a congested state in which the average time-headway is
less than a critical value, the typical sketch of total delay undergoes
an abrupt change which is shown in
the following set of figures.  Unexpectedly, there are two minima for
delay which are symmetrically located from each other. The graphs tell us
that in order to minimise the total delay we have to choose one of the 
minima. In other words, we have to trample the right of
one direction in
favour of the other direction. In the context of statistical physics, this
is
equivalent to breaking the symmetry of the problem. In practice, we can
 periodically change the
trampled road
from A to B. This phenomenon is rather analogous to the spontaneous
symmetry
breaking in equilibrium statistical mechanics. Creation of two
valleys in the delay curve is reminiscent of double-valley creation in
the
free energy function. In critical phenomena, the physical system will
ultimately be found in one of the valleys. In the traffic context, we
should
signalizes the lights in such a way that-despite the equal conditions for
both
streets- the majority of green time be
allocated to one of streets. Quantitatively, increasing the congestion of
the
entering cars, leads to a broader height of the barrier. It should be
mentioned that the valleys are each discontinuously separated from the
barrier.
If we decrease the entrance rate, the barrier shrinks to a smaller size
and finally at a critical value , it disappears.

\begin{figure}\label{Fig3}
\epsfxsize=5.5truecm
\centerline{\epsfbox{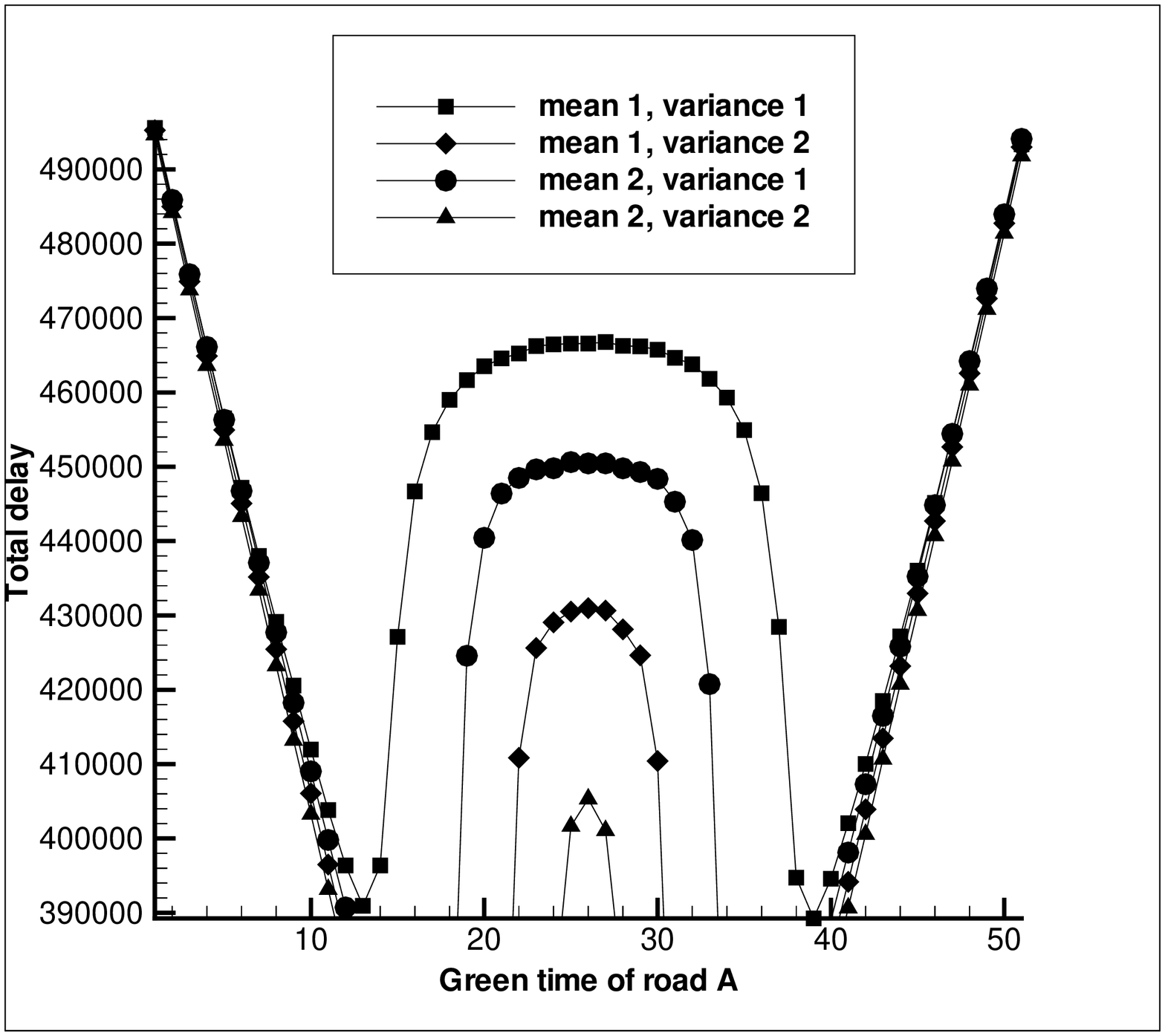}}
\end{figure}
\vspace{0.05 cm}
{\small{Fig.~3: congested traffic state. The delay curve has two minima
and choosing one of them leads to breaking the symmetry between the two
roads.
The more congested the flow is, the broader the barrier. The system has
evolved for 5000 time-steps and T is set to 50. } }

\begin{figure}\label{Fig4}
\epsfxsize=5.5truecm
\centerline{\epsfbox{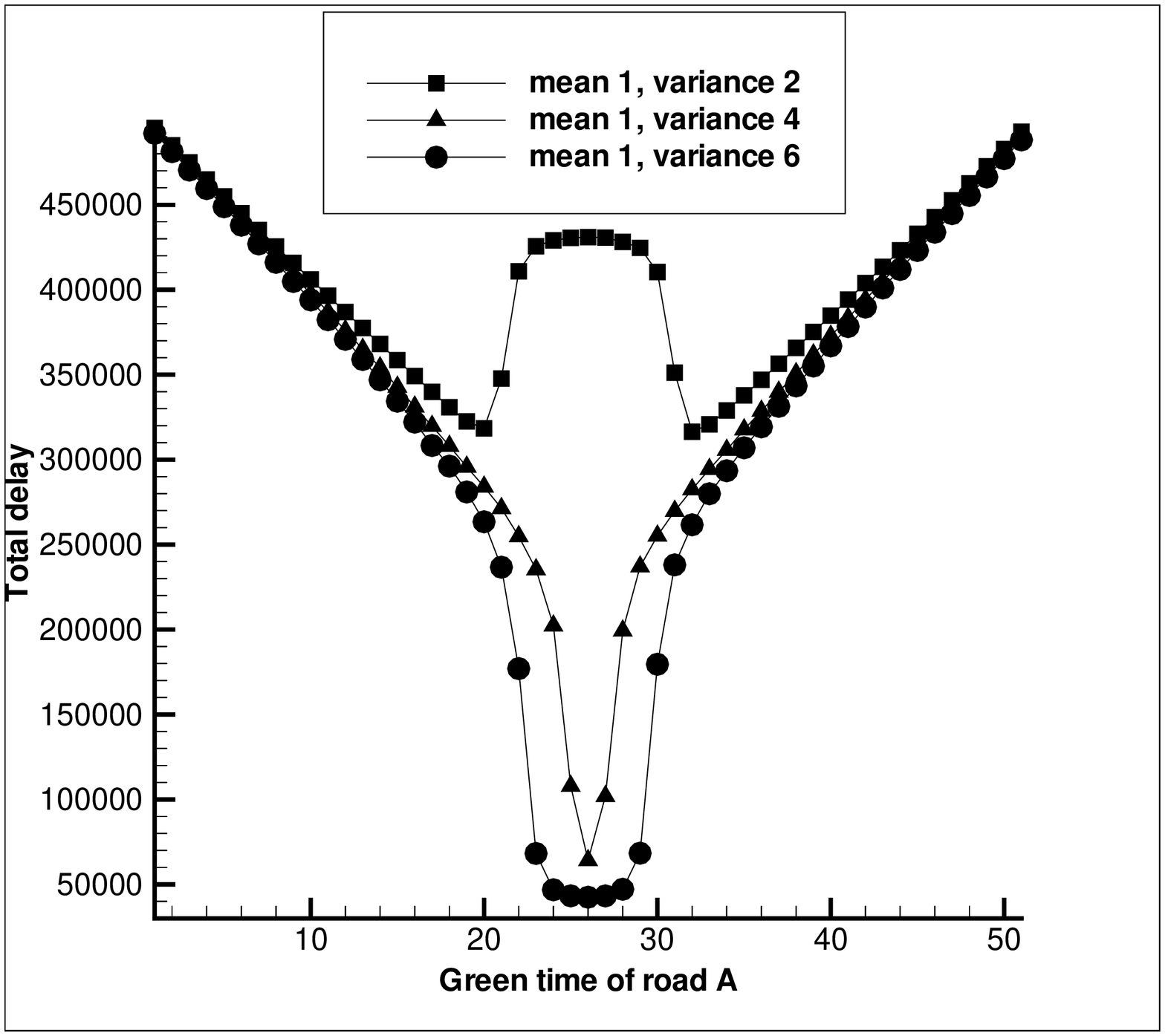}}
\end{figure}
\vspace{0.05 cm}
{\small{Fig.~4: for fixed value of space gap average, there is a critical
variance above which the phase transition between the light and congested
state is manifest. For average=1, the critical variance is 4. The system
has evolved for 5000 time-steps and T is set to 50.} }

\begin{figure}\label{Fig5}
\epsfxsize=5.5truecm
\centerline{\epsfbox{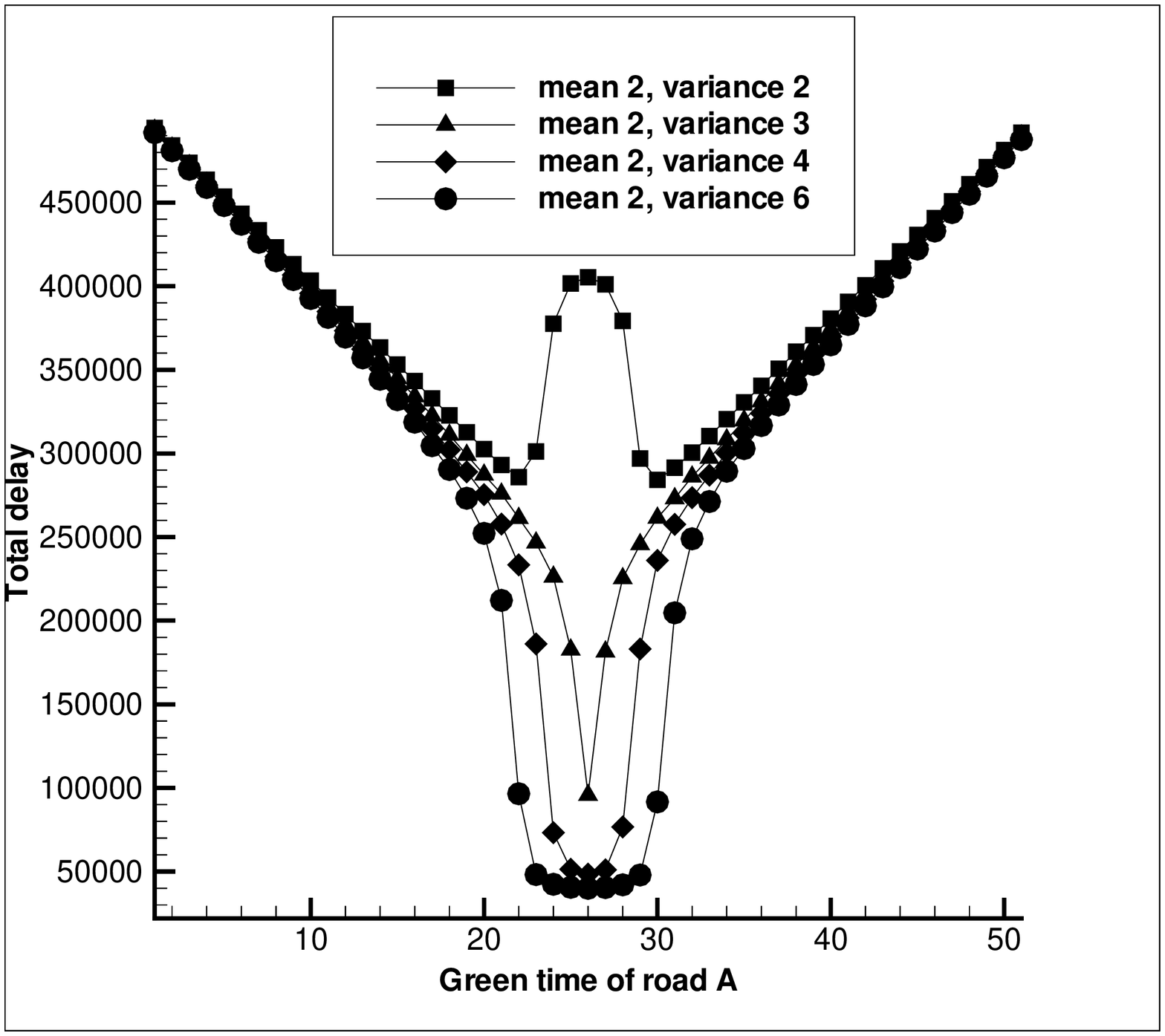}}
\end{figure}
\vspace{0.05 cm}
{\small{Fig.~5: the effect of increasing the variance for
average=2 is shown. The critical variance is 3. the system has evolved
for 5000 time-steps and T is 50. } }

\subsection{ Asymmetric case }

In the asymmetric states, the entering rates into the roads differ from
each
other. However, two distinct phases are identified as before. In cases
where both roads are carrying light flows, the asymmetry between entering
rates causes a displacement on the position of flat optimum region but
does not change the overall characteristics of the delay curve. The
following figures shows the effect of asymmetry on the delay curve. An
interesting asymmetric state is the intersection of a major to a minor
street. A large fraction of urban crossroads lie in the category of 
major-to-minor. The signalisation of these types of crossing is still a
controversial subject. The main reason is that in most of intelligent
real-time controller systems, the signalisation of these crossroads are
highly affected by the major crossroads signalisation schemes
which frequently overlook the local optimisation of minor crossroads. In
our model, a minor crossing is modelled by a light traffic in one road and
a congested one in the other road. The following graphs depicts the
behaviour of the delay curve.  

\begin{figure}\label{Fig6}
\epsfxsize=5.5truecm
\centerline{\epsfbox{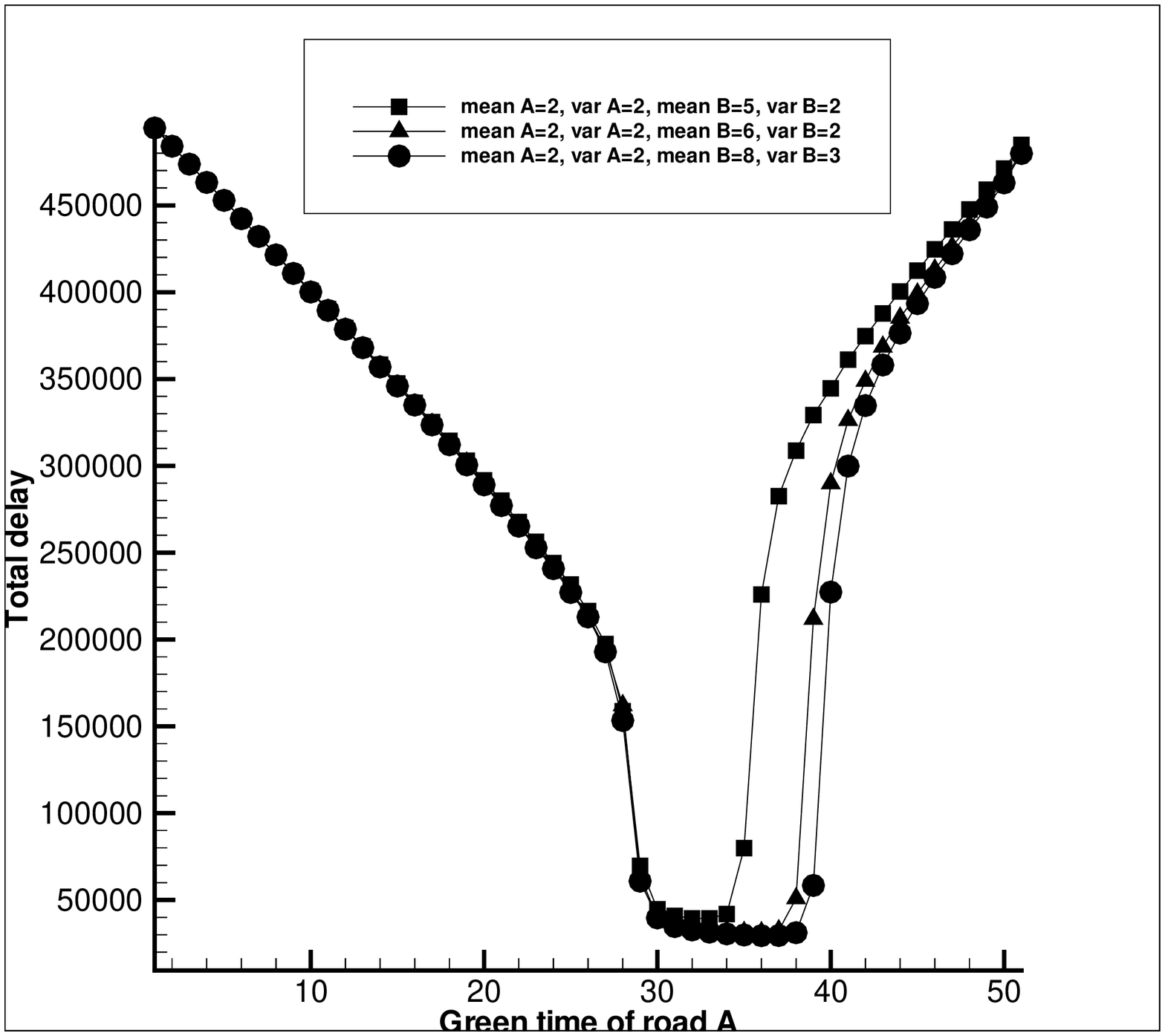}}
\end{figure}
\vspace{0.05 cm}
{\small{Fig.~6: The asymmetry between roads displaces the flat region
toward the road with lighter traffic. } }

\section{one-way to two-way crossroads}

In this section we examine the traffic state at a crossroads where one
street is one-way while the other one conducts a two-way flow. We take
the
traffic flow on the eastward street to be one-way.  
In one the most popular signalisation schemes, the total cycle period is
divided into three movement phases. During the first phase which lasts
for
$T_{1}$ seconds, the traffic on one of roads can move while the
lights are red for the other two roads. In the second phase which lasts
from $T_{1}$ until $T_{2}$ the second road's light switches to green and
the other two traffic lights are kept red and finally from $T_{2}$ until
$T$ only the third road is allowed to move. The simulation rules are the
same
as one-way to one-way rules. The total delay is a function of $T,T_1,T_2$
as well as inflow rates. The following three dimensional graphs show the
behaviour of delay curve as a function of $T_1$ and $T_2$. In accordance
with the results of previous sections, two distinct states are
already identified.
In the symmetric state, which corresponds to an overall lightness
of the traffic volume, the flat optimal region is now extended to a two
dimensional
area. The boundary of this area is sharply separated from a highly valued
delay area of the delay space. The results warn us to take care in tuning
the signalisation times $T_1$ and $T_2$ in such a way that we remain
inside the flat region. The other state, corresponding to a
symmetry broken state, shows the behaviour of the delay curve in congested
cases. In
the symmetry broken states, three minima valleys are formed and in order
to minimise the total delay, one should break the symmetry in favour of
one chosen direction. Analogous to one-way crossroads, the valleys are
discontinuously separated from each other by a two dimensional barrier.

\begin{figure}
\centerline{\mbox{\psfig{figure=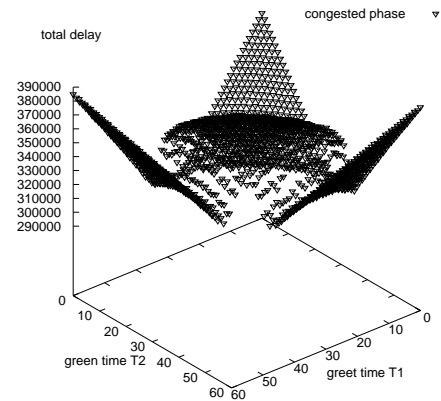, angle=270, width=6.3 cm}}}
\caption{total delay as a function of green times $T_1$ and $T_2$ in a
totally symmetric state. the
system has evolved for 2000 time-steps and the value of $T$ has been set
to 60. the average as well as the standard deviation values of arrival
rates distribution functions are two.} \label{ebram
gholi}
\end{figure}

\begin{figure}
\centerline{\mbox{\psfig{figure=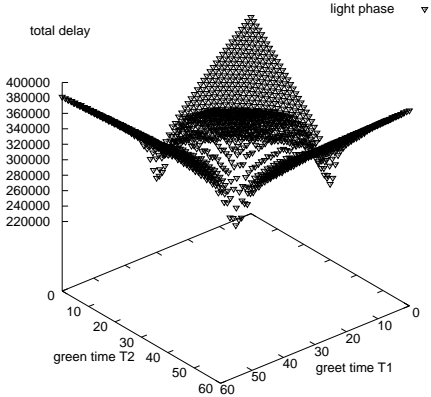, angle=270, width=6.3 cm}}}
\caption{total delay as a function of green times $T_1$ and $T_2$ in a
totally symmetric state. the
system has evolved for 2000 time-steps and the value of $T$ has been set
to 60. the averages are two while the standard deviation of
the arrival distribution functions are three. it can be seen that
analogous to the one-way to one-way intersection, for
average=2, the critical variance is three. } \label{ebram
gholi}
\end{figure}

\begin{figure}
\centerline{\mbox{\psfig{figure=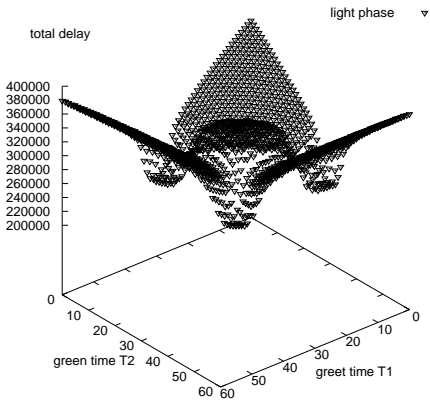, angle=270, width=6.3 cm}}}
\caption{total delay as a function of green times $T_1$ and $T_2$ in a
totally symmetric state. the
system has evolved for 2000 time-steps and the value of $T$ has been set
to 60. the averages are two while the standard deviation of
the arrival distribution functions are four. } \label{ebram
gholi}
\end{figure}

\section{Summary and concluding remarks}

Single crossroads are
not just an artificial realization of city network. Marginal urban 
areas are places where single crossroads are frequently designed 
and operated. Apart from this convincing fact, it has always been a
subject of argument whether to control a crossroads under a centralized
or decentralized scheme. In special circumstances, decentralized local
adaptive strategies operate more effectively than globally adaptive
strategies \cite{porche,huberman,brockfeld} and often show a very good
performance.
Therefore, investigations on single junctions can be of practical
relevance for various applications in city traffic. For this purpose,
we have developed and analysed a prescription for the traffic light
signalisation at a single crossroads. Our investigation predicts the
existence of traffic states in which to
gain the maximum
throughput traffic, one manifestly has to create inequality in favour
of
one of the flow directions. The phenomenon rather resembles the
breaking of the symmetry in the onset of phase transitions in equilibrium
statistical physics.\\
Once the conflicting
movement directions are enhanced over two, i.e., one-way to two-way
crossing, another question
arises which governs the designing the movement structure of the
crossroads itself. It can easily be shown that in the case of one-way to
two-way crossroads, five different types of signalisation can be defined
each of which corresponds to a distinct movement structure. 
In this paper we have discussed only one of them. Another frequent example
is parallel moving of north-bound together with east-bound vehicles toward
north in one phase, parallel moving of south-bound together with
east-bound vehicles toward south in the second phase and finally the
remaining non-conflicting directions in the third phase. Prior
to optimisation of green times, one first should determine the optimal
movement structure and then optimise the signalisation of the optimal
structure. The detailed analysis of the movement structure in one-way to
two-way and more generally the intersection of a couple two-way roads will
be discussed in a separate paper.\\

To conclude, we have proposed an
optimising adaptive decentralized scheme
for a single crossroads on the basis of {\it minimised total delay }
concept. Our next objective is to optimise the traffic flow in a simple
urban
structure of junctions e.g. a small-sized cluster of crossroads. The
results will be reported elsewhere.

\section{ Acknowledgments}
 
We are grateful to {\it Tehran Traffic
Control Company }. We wish to thank Reza Asgari and Bahman Davoudi
for graphical helps and would like to express our gratitude to R.
Sorfleet, for reading the manuscript.

\bibliographystyle{unsrt}

\begin{thebibliography}{99}



\bibitem{css99}
D. Chowdhury, L. Santen and A. Schadschneider,
{\em Physics Reports}, {\bf 329}, 199 (2000).



\bibitem{helbbook} D.\ Helbing, {\em Vehrkersdynamik: Neue Physikalische
Modellierungskonzepte}, Springer, Berlin,1997;  {\em Traffic and related
self-driven many particle systems }, to appear in Rev. of Mod. Physics.





%\bibitem{tgf95} D.E.\ Wolf, M.\ Schreckenberg and A.\ Bachem (eds) {\em
%Traffic and granular flow} (World Scientific, Singapore, 1996).



\bibitem{tgf97} D.E\ Wolf and M.\ Schreckenberg (eds.) {\em Traffic and
granular flow} (Springer, Singapore, 1998).



\bibitem{tgf99} \ H.J.\ Herrmann, D.\ Helbing, M.\
Schreckenberg and D.E.\ Wolf (eds.) {\em Traffic and Granular flow}
 (Springer, Berlin, 2000).
    


\bibitem{prigogine} I. Prigogine and R. Herman, {\em Kinetic theory of
vehicular traffic}, (Elsevier, Amsterdam, 1971).


\bibitem{bml} O.\ Biham, A.\ Middleton and D.\ Levine, Phys.Rev. {\bf A
}{\bf 46}, R6124 (1992).




\bibitem{ns92} K. Nagel, M. Schreckenberg, {\em J.Phys.I France} {\bf 2},
2221
(1992).




\bibitem{chung} K.H. Chung, P.M. Hui and G.Q. Gu, {\em Phys. Rev. E} {\bf
51}, 772 (1995).


\bibitem{cuesta} J.A. Cuesta, F.C. Martines, J.M. Molera and A. Sanchez,
{\em Phys. Rev. E} {\bf 48}, R4175 (1993).



\bibitem{nagatani} T. Nagatani, {\em J. Phys. Soc. Japan} {\bf 63}, 1228
(1994).




\bibitem{torok} J. T\"or\"ok and J. Kertesz, {\em Physica A} {\bf 231},
515 (1996).




\bibitem{cs} D.\ Chowdhury and A.\ Schadschneider, {\em Phys. Rev. 
E} {\bf 59 },
R 1311 (1999).



\bibitem{brockfeld} E.\ Brockfeld, R.\ Barlovic, A.\ Schadschneider and  
and M.\ Schreckenberg, preprint, cond-mat/0107056





\bibitem{foolad} M.E. Fouladvand and M. Nematollahi, to appear in European
Phys. J. B (2001), cond-mat/0106268



%\bibitem{privbook} {\em Non-Equilibrium Statistical Mechanics in One
%Dimension}, edited by V. Privman (Cambridge University Press, Cambridge,
%England, 1997).




%\bibitem{book}
% {\em Traffic Flow Fundamentals}, Prentice Hall (1990) by A.D.\
%May. {\em Transportation and Traffic Theory}, Elsevier (1993) by C.F
%\ Daganzo. 




\bibitem{robertson} D.I. Robertson and R.D. Bretherton, {\it Optimizing
networks of traffic signals in real-time: the SCOOT method }, {\em IEEE
Transportation on Vehicular Technology}, {\bf 40}, 11 (1991).







\bibitem{porche}I. Porche, M. Sampath, Y.-L. Chen, R. Sengupta and S.
Lafortune, {\em A decentralized scheme for real-time optimization of
traffic signal } in the proceeding of 1996 {\it IEEE International
Conference on Control Applications}, 582-589.


\bibitem{huberman} B. Faieta and B.A. Huberman, {\it firefly : A
synchronisation strategy  for urban traffic control}, Xerox Palo alto
Research Centre, Palo Alto, CA 94304.




%\bibitem{nagatani2} T. Nagatani, {\em J. Phys. Soc. Japan} {\bf 63}, 1228
%(1994). 






%\bibitem{horiguchi} T. Horiguchi and T. Sakakibara, {\em Physica A} {\bf
%252}, 388 (1998); {\em Interdisc. Inf. Sci.} {\bf 4}, 39 (1998). 



%\bibitem{freund} J. Freund and T. P\"oschel, {\em Physica A} {\bf 219},
%95
%(1995).



%\bibitem{chopard} B. Chopard, P.O Luthi and P.A. Queloz, {\em J. Phys. A}
%{\bf 29}, 2325 (1996).







%\bibitem{ito} M. Schreckenberg, A. Schadschneider, K. Nagel and N. Ito,
%{\em Phys. Rev. E} {\bf 51}, 2939 (1995).



%\bibitem{brockfeld} Elmar Brockfeld, {\em Simulation von Stadtverkehr
%mittels Zellularautomaten} (Diplomarbeit), University of Osnabr\"uck,
%2000.



%\bibitem{scats1} P. Lowrie, {\it SCATS: A Traffic Responsive Method for
%Controlling Urban Traffic}, tech. rep., {\em Road and Traffic Authority},
%NSW, Australia.


%\bibitem{scats2} A.G. Sims, "SCATS: the Sydney co-ordinated adaptive
%system", in the proceeding of the {\it Engineering Foundation Conference
%on Research Priorities in Computer Control of Urban Traffic Systems}, 12,
%1979.
 

\end{thebibliography}

\end{document}